\documentclass[sigconf,authorversion]{acmart}

\usepackage{url}
\usepackage{graphicx}
\usepackage{booktabs}
\usepackage{tabularx}
\usepackage{booktabs}
\usepackage{balance}

\begin{document}

\title{A Demand-Side Viewpoint to Software \\ Vulnerabilities in WordPress Plugins}

\author{Jukka Ruohonen}
\affiliation{\institution{University of Turku, Finland}}
\email{juanruo@utu.fi}

\begin{abstract}
WordPress has long been the most popular content management system (CMS). This CMS powers millions and millions of websites. Although WordPress has had a particularly bad track record in terms of security, in recent years many of the well-known security risks have transmuted from the core WordPress to the numerous plugins and themes written for the CMS. Given this background, the paper analyzes known software vulnerabilities discovered from WordPress plugins. A demand-side viewpoint was used to motivate the analysis; the basic hypothesis is that plugins with large installation bases have been affected by multiple vulnerabilities. As the hypothesis also holds according to the empirical results, the paper contributes to the recent discussion about common security folklore. A few general insights are also provided about the relation between software vulnerabilities and software maintenance.
\end{abstract}

\copyrightyear{2019}
\acmYear{2019}
\setcopyright{acmlicensed}
\acmConference[EASE '19]{Evaluation and Assessment in Software Engineering}{April 15--17, 2019}{Copenhagen, Denmark}
%\acmBooktitle{Evaluation and Assessment in Software Engineering (EASE '19), April 15--17, 2019, Copenhagen, Denmark}
%\acmPrice{15.00}
\acmDOI{10.1145/3319008.3319029}
%\acmISBN{978-1-4503-7145-2/19/04}

% Generated at http://dl.acm.org/ccs.cfm
%



\maketitle

\section{Introduction}

The WordPress content management system is undoubtedly one of the great success stories of open source software (OSS). This CMS written in the PHP programming language has long been the most popular CMS worldwide. In fact, it has been estimated that even as much as one-third of all websites would be powered by WordPress~\cite{Cabot18}. But with great success comes great responsibility~\cite{MansfieldDevine15}. When it comes to security, WordPress has been a sorrowful representative of OSS. Vulnerabilities are frequently discovered from the CMS, and mass-scale compromises are commonly reported in media~\cite{Threatpost19a}. By no means is WordPress alone making these headlines, however. Many websites use outdated and deprecated releases of the PHP language \cite{Ruohonen16WIMS, Ruohonen17APSEC}, for instance. All this said, in recent years particularly the management of security issues has greatly improved in the WordPress ecosystem~\cite{Cabot18}. The ecosystem concept is also useful for framing this study against existing research.

Recent research has made good progress on understanding vulnerabilities in software ecosystems through analyzing ``hard'' library dependencies~\cite{ZapataKula18, Vaidya19}. Despite of these advances, library dependencies paint only a limited picture on whole software ecosystems. In the WordPress ecosystem particularly important are the numerous plugins and themes written for the CMS. It is presumably also these complementary software elements that nowadays pose the greatest security risks for WordPress deployments~\cite{Threatpost19a, Trunde15}. Even though plugins are reviewed by a WordPress team prior to submission into the official hosting portal~\cite{Cabot18}, new plugin vulnerabilities are discovered on day-to-day basis. In fact, some practitioners have contemplated that even ninety-nine out of a hundred WordPress plugins could be vulnerable~\cite{MansfieldDevine15}. The already discovered and publicly disclosed plugin vulnerabilities are the topic of this study. To motivate the topic and the analysis, a specific demand-side viewpoint is pursued.

The background relates to counterintuitive findings about common security folklore. In particular, it has been observed that up-to-date WordPress deployments with large user bases are a frequent target for exploitation, although a common folk wisdom would tell the opposite~\cite{Moore16}. Though, it should be remarked that the existing empirical evidence is not entirely unequivocal. For instance, the popularity of websites has been observed to correlate with the adoption of basic web-related security features~\cite{vanGoethem14}. Likewise, less popular websites that are known to have been vulnerable to cross-site scripting (XSS) have been observed to use these security features less frequently~\cite{Ruohonen18SQAMIA}. Despite of these observations, the real contribution from the counterintuitive findings stems from the ways to think about common security folklore and the subsequent need for evidence-based research~\cite{Moore16}. One way to think about known vulnerabilities is to think about supply and demand. 

If the supply-side factors include things like the availability of static analysis tools \cite{Nunes18} and the ease of searching and fingerprinting WordPress deployments~\cite{ZhangYang12}, popularity would be a notable demand-side factor. Accordingly, there should be only a small incentive to find new vulnerabilities from unpopular plugins. To examine such an incentive indirectly, the research question (RQ) examined is simple: do large installation amounts increase the amount of WordPress plugin vulnerabilities discovered and disclosed? Given this research question, the structure of the paper's remainder is straightforward: the dataset examined is elaborated in Section~\ref{sec: data}, the empirical results are presented in Section~\ref{sec: results}, and a discussion about the findings, limitations, and future directions follows in Section~\ref{sec: discussion}.

\clearpage
\pagebreak
\section{Data}\label{sec: data}

The dataset was assembled from the following three sources:

\begin{enumerate}
\item{The primary data source is the so-called WPScan Vulnerability Database (WPVDB) \cite{WPVDB18a}. In contrast to many other vulnerability databases, WPVDB is a specialized database exclusively targeting the core WordPress as well as the numerous third-party plugins and themes written for the popular CMS. Furthermore, WPVDB is a rather unique in the sense that the primary rationale for the database is to explicitly supply data for the associated WPScan, a black-box vulnerability scanner for online WordPress deployments.}
\item{The second data source is the official online portal for hosting WordPress plugins \cite{WordPress18a}. This portal provides the necessities for plugin development, including version control system hosting and forums for user feedback. For each plugin listed in WPVDB, the portal's online interface was queried for retrieving meta-data about the plugin. If a plugin could not be mapped from WPVDB to the online portal, it is excluded from the dataset and the forthcoming empirical analysis.}
\item{The third and final source is the conventional National Vulnerability Database (NVD) \cite{NVD18b}. If a given plugin vulnerability archived to WPVDB was accompanied with an identifier for Common Vulnerabilities and Exposures (CVEs), this identifier was used to retrieve further data from NVD. Although WPVDB provides additional meta-data for some vulnerabilities, the scope of this data is limited and not all plugin vulnerabilities are covered. Therefore, the auxiliary data from NVD provides a more robust basis for a few descriptive but important insights about the plugin vulnerabilities.}
\end{enumerate}

A further point should be made about abstractions. Each plugin in WPVDB may be affected by multiple vulnerabilities, a single vulnerability entry in WPVDB may reference multiple distinct CVEs, and a single unique CVE may reference multiple entries in WPVDB. These abstraction inconsistencies are typical to practical tracking and archiving of software vulnerabilities~\cite{Doyle11}. For instance, vendors oftentimes aggregate fixes for multiple CVE-referenced vulnerabilities into a unified patch set, which is typically further abstracted into a single security advisory delivered to users and system administrators. While there is thus no single right way to abstract and count vulnerabilities, the abstraction choices have direct consequences for empirical analysis. Because in this paper the amount of installations is the primary independent metric of interest, the only sensible way is to perform the empirical analysis at the plugin-level. Thus, the units of analysis are WordPress plugins that have been affected by one or more vulnerabilities, as counted in WPVDB. In addition, a few descriptive observations are delivered through the CVE-level by using NVD's abstraction for counting vulnerabilities.

\section{Results}\label{sec: results}

\subsection{Overview}

\subsubsection{Sample characteristics}

\begin{figure}[th!b]
\centering
\includegraphics[width=\linewidth, height=4.0cm]{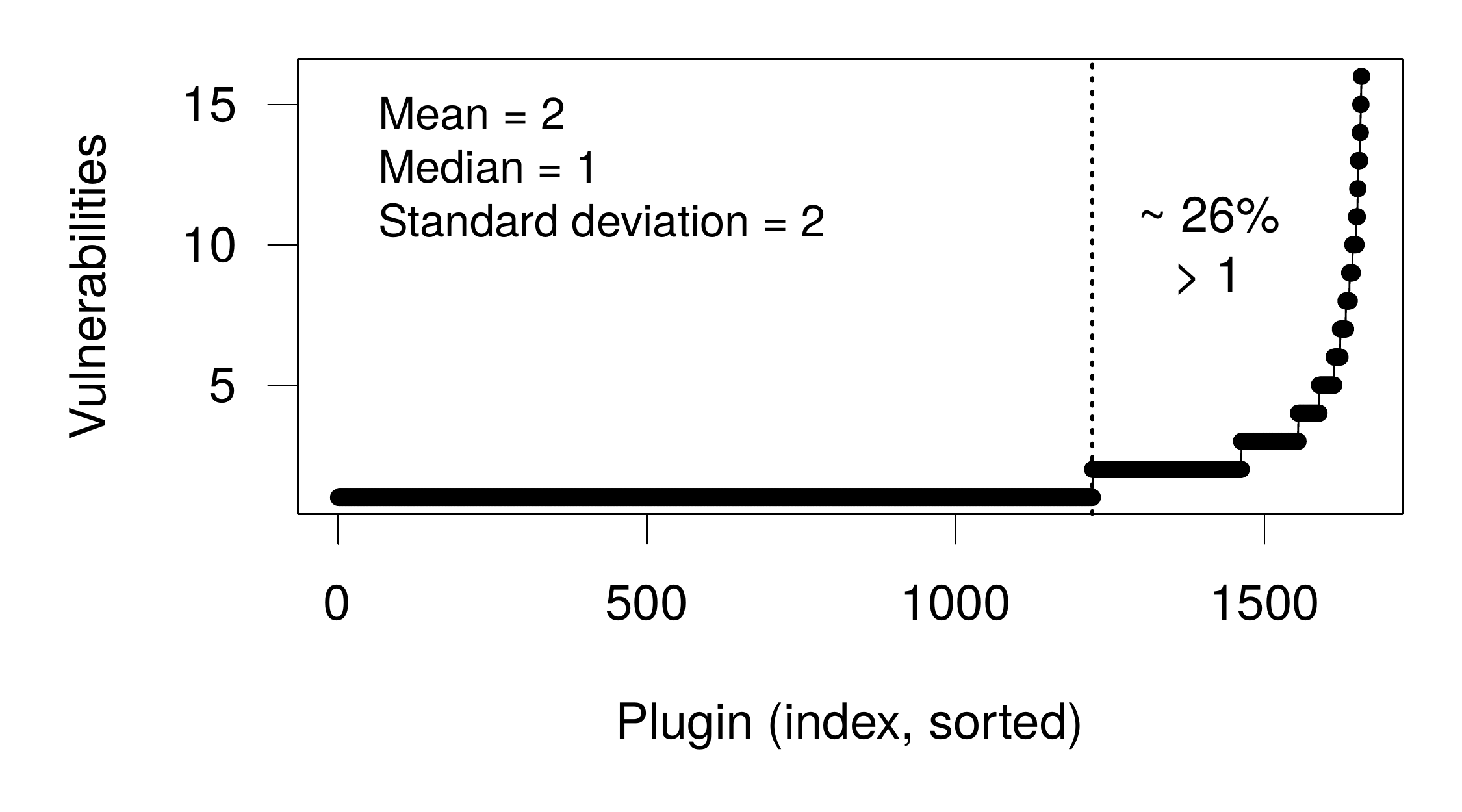}
\caption{Vulnerability Counts (WPVDB's abstraction)}
\label{fig: counts}
\end{figure}

The empirical dataset assembled contains $1,657$ plugins that were affected by $2,629$ vulnerabilities according to WPVDB's abstraction for counting. These numbers are sufficient for a couple of preliminary points about the folk wisdom examined. The first point is that not many plugins have been vulnerable---according to the online portal~\cite{WordPress18a}, there were over fifty-five thousand WordPress plugins available for download at the time of data collection. Thus, according to the dataset, roughly only about five percent of these plugins have been vulnerable at some point in time. Of course, it is difficult to assess the reliability of this observation; many of the plugins have presumably never been audited, and, hence, numerous existing vulnerabilities likely remain undiscovered and undisclosed. Nevertheless, $\sim 5\%$ is such a small value that it seems reasonable to recommend avoiding words such as ``most'' or ``majority'' when discussing about vulnerable WordPress plugins. The second point is that only about 26\% of the plugins observed have been affected by multiple vulnerabilities. As has been observed also previously~\cite{Koskinen12, Walden10b}, the distribution across the plugins is highly skewed, however. A few plugins have been affected by many vulnerabilities (see Fig.~\ref{fig: counts}). The cases with multiple vulnerabilities are the main interest in the forthcoming regression analysis. Before continuing to formal statistical analysis, the CVE-level counting can be used for a few interesting observations.

\subsubsection{CVEs}\label{subsec: CVEs}

Only about 28\% of the plugin vulnerabilities in the sample are accompanied with one or more CVEs that have valid entries in NVD. Although this amount is quite small, it is fairly typical for specialized vulnerability databases targeting small open source projects for which CVEs may not be always allocated~\cite{Ruohonen18IWESEP}. It seems also reasonable to assume that the few forthcoming CVE-based descriptive observations generalize to all plugin vulnerabilities in the database due to the rather generic nature of these observations.

\subsubsection{NVD}

\begin{figure}[th!b]
\centering
\includegraphics[width=\linewidth, height=4.2cm]{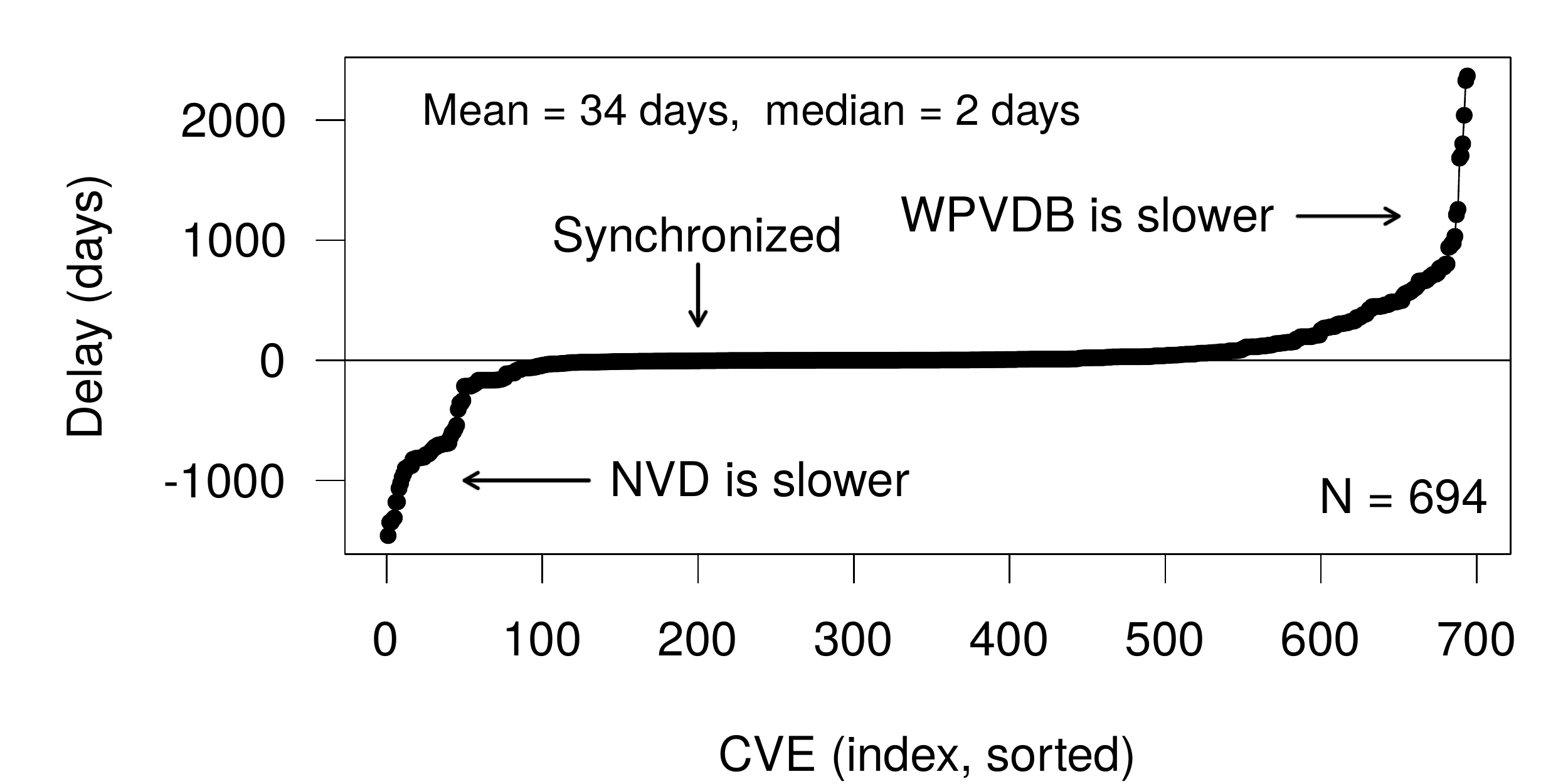}
\caption{Publication Delays Between NVD and WPVDB}
\label{fig: years}
\end{figure}

The first observation can be made from Fig.~\ref{fig: years}, which visualizes the time delays between the CVE-referenced publication dates in WPVDB and NVD, using the earliest dates (the smallest timestamps) for the former in case multiple CVEs are present. Because most of the values are zero, the two databases appear to be implicitly synchronized with each other; a plugin vulnerability appearing in NVD tends to appear during the same day in WPVDB, or the other way around. That said, there is also a sizable amount of positive values, meaning that many of the plugin vulnerabilities were archived to NVD before these appeared in WPVDB. The slightly smaller amount of negative values is also interesting because these cases implicitly justify the use of WPVDB's data for monitoring online deployments. The reason why NVD is sometimes slower may relate to the online sources monitored by WPVDB's maintainers for gaining information about plugin vulnerabilities.

\subsubsection{Information sources}

\begin{figure}[th!b]
\centering
\includegraphics[width=\linewidth, height=8.2cm]{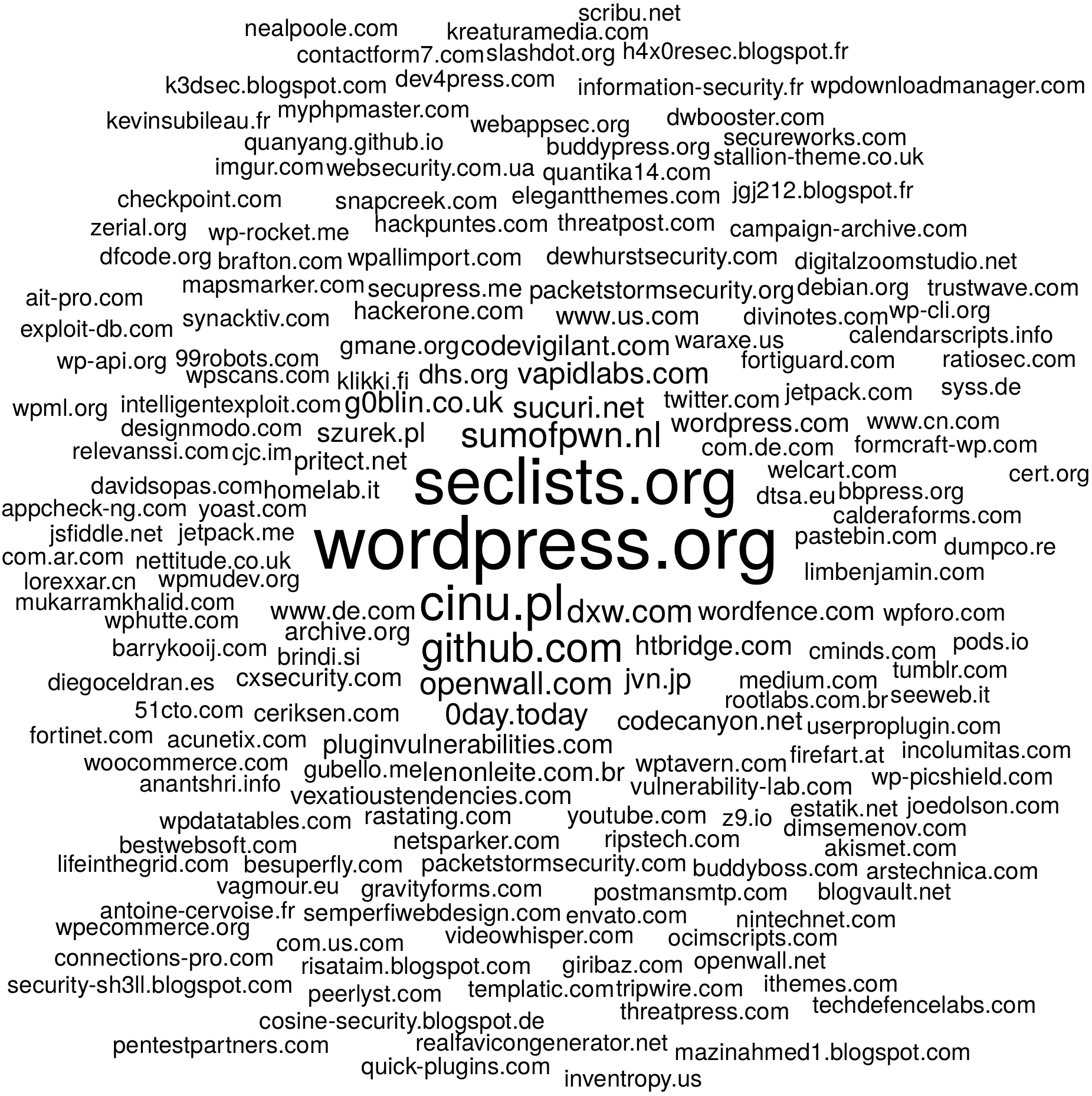}
\caption{Domain Names (2LD-TLDs) of the Reference URLs}
\label{fig: domains}
\end{figure}

Like many \cite{Ruohonen18IST, ZapataKula18}, but not all~\cite{Ruohonen18IWESEP}, vulnerability databases, WPVDB provides hyperlinks to the original information sources. To illustrate the main sources, Fig.~\ref{fig: domains} visualizes the second-level (2LD) and top-level (TLD) domain names extracted from the uniform resource locators (URLs) present in the hyperlinks, using the so-called public suffix list for the comparisons~\cite{SuffixList19}. Although social media has been suspected to play an increasingly important role~\cite{Sauerwein18, ZongRitter19}, the illustration clearly indicates that plugin vulnerabilities are commonly disseminated through very traditional channels for communicating security issues in the OSS context. In fact, social media is hardly even present. The most frequent domain name (2LD-TLD) is \texttt{wordpress.org}, which hosts the plugin portal itself, including the wikis, version control systems, and bug trackers often used for WordPress plugin development. The second most frequent domain name is \texttt{seclists.org}, which hosts and archives a number of security mailing lists. A closer look reveals also numerous diverse information sources. The examples include hosting services, blogs, company websites, bug trackers, bug bounty platforms, so-called pastebins, other databases, media outlets, online archives, and personal homepages. Taken together, these diverse sources are a good example on the practical challenges for current vulnerability tracking~\cite{Ruohonen18IST}. Such practical challenges  also translate into research challenges: mining a single software repository (or a few repositories, for that matter) is assuredly inadequate for empirically observing most \textit{known} software (plugin) vulnerabilities.

\begin{figure}[th!b]
\centering
\includegraphics[width=\linewidth, height=7.9cm]{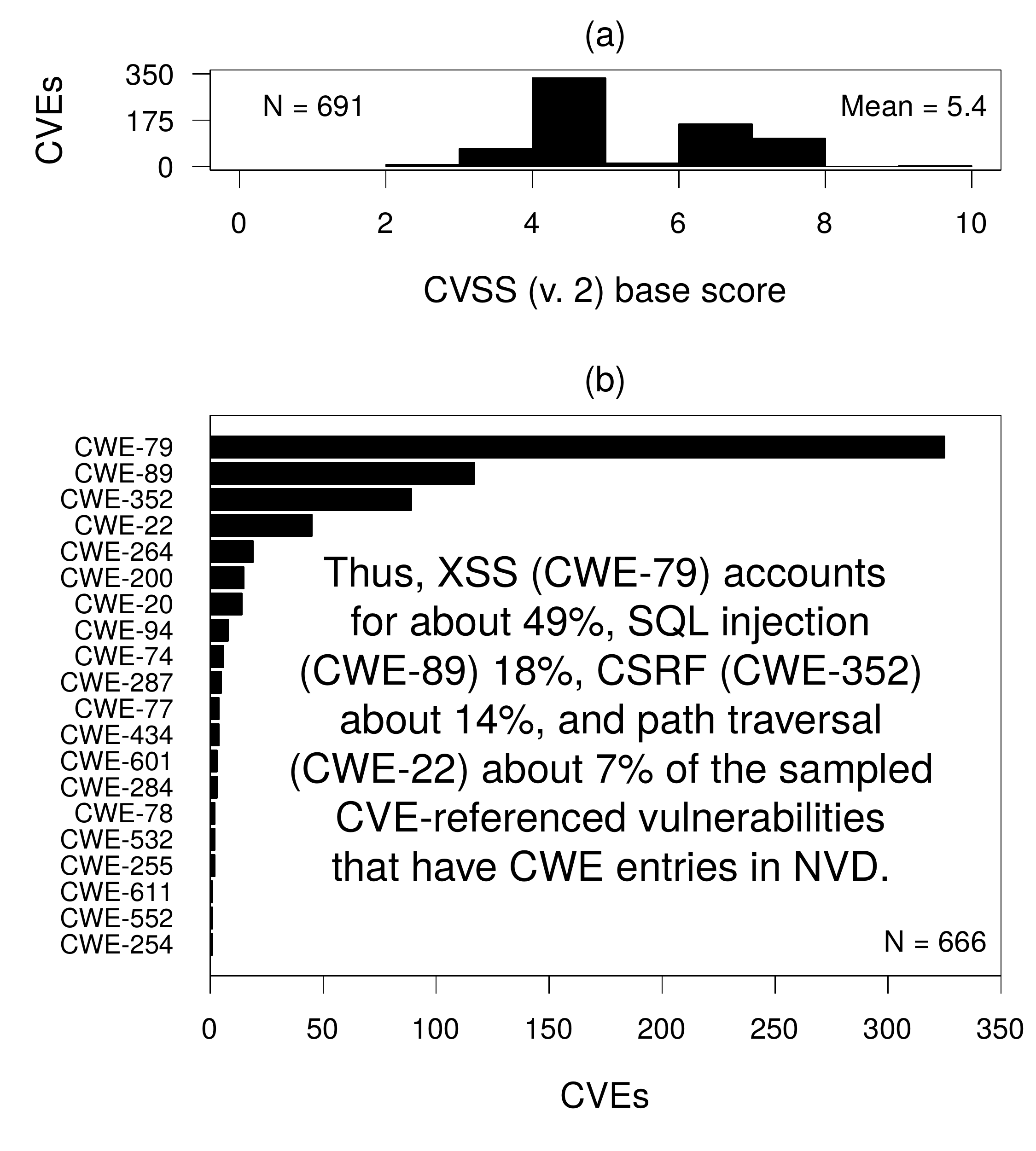}
\caption{Severity (CVSS) and Weaknesses (CWE)}
\label{fig: cvss cwe}
\end{figure}

\subsubsection{Severity and weaknesses}

In addition to reliability and validity compared to WPVDB, the benefits of NVD include additional information about vulnerabilities. In particular, the Common Vulnerability Scoring System (CVSS) and the Common Weakness Enumeration (CWE) framework provide information about the severity of the vulnerabilities and the typical weaknesses behind these. As is well-known~\cite{Ruohonen18IWESEP, Santos17, ZhangMalhotra18}, it should be remarked that not all of the CVEs observed have CVSS and CWE entries in NVD due to delays and other database maintenance issues. In any case, the results regarding these frameworks are summarized in Fig.~\ref{fig: cvss cwe}. The so-called base CVSS (v.~2) scores shown in the plot (a) indicate only modest severity. The median base score is five. This average score is largely explained by the substantial amount of XSS vulnerabilities. As can be seen from the plot (b), cross-site scripting, structured query language (SQL) injection, and cross-site request forgery (CSRF) account for over 80\% of the weaknesses behind the CVE-referenced vulnerabilities in the sample. This CWE-based ranking conforms with existing results~\cite{Mesa18}. These observations also agree with existing results about typical weaknesses in typical PHP applications~\cite{Huynh10, Walden10a}. In general, the plugin vulnerabilities observed are supposedly not that different from the vulnerabilities that have affected the core WordPress itself~\cite{Walden10b}. Another point is that only CWE-434 (unrestricted file uploads) seems specific to PHP~\cite{MITRE19a}. Thus, all in all, it could be also said that the plugin vulnerabilities observed are typical to web applications in general.

\subsection{Meta-Data}\label{subsec: meta-data}

\subsubsection{Installations}

\begin{figure}[th!b]
\centering
\includegraphics[width=\linewidth, height=9.5cm]{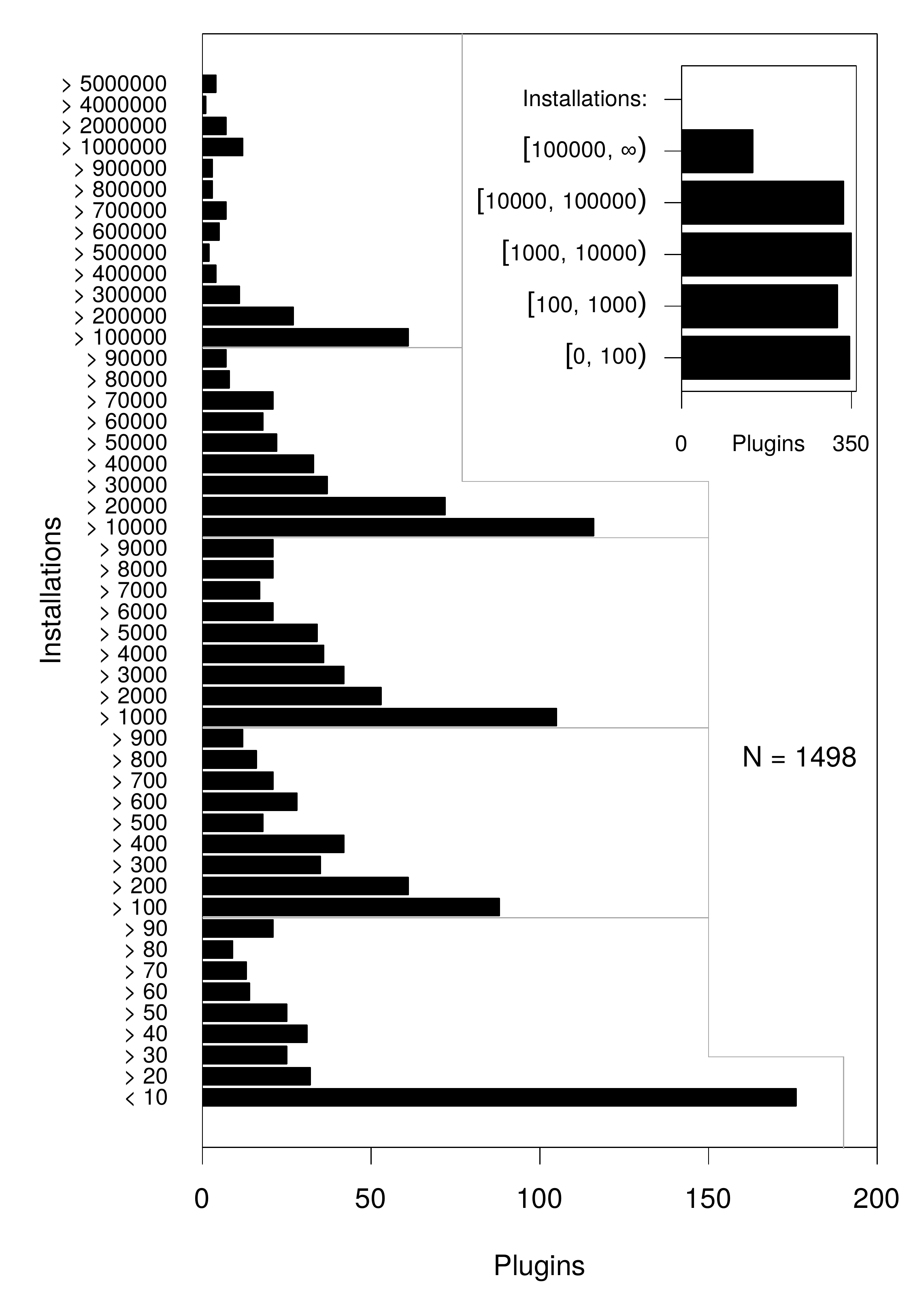}
\caption{Approximate Number of Installations}
\label{fig: installations}
\end{figure}

The few remaining descriptive observations are based on the meta-data scraped from the WordPress online portal. The first observation relates to the approximate installation amounts. Although WordPress has received attention in Internet measurement research~\cite{vanGoethem14, VasekMoore14}, there is no good understanding on how many websites are actually powered by the CMS, let alone on how many of these online deployments are running with plugins. While keeping this point in mind, the outer plot in Fig.~\ref{fig: installations} displays the approximations given by the maintainers of the online portal (the $159$ missing observations refer to deprecated plugins for which meta-data is not necessarily provided). The range is wide: there are many plugins with approximately less than ten online installations and a few plugins that have been installed in over one hundred thousand WordPress deployments. The crudeness of these meta-data approximations is reflected in the spikes around the powers of ten. Re-coding is therefore justified---the inner plot in the figure shows the re-coded 5-fold variable used in the regression analysis.

\begin{figure}[th!b]
\centering
\includegraphics[width=\linewidth, height=6.3cm]{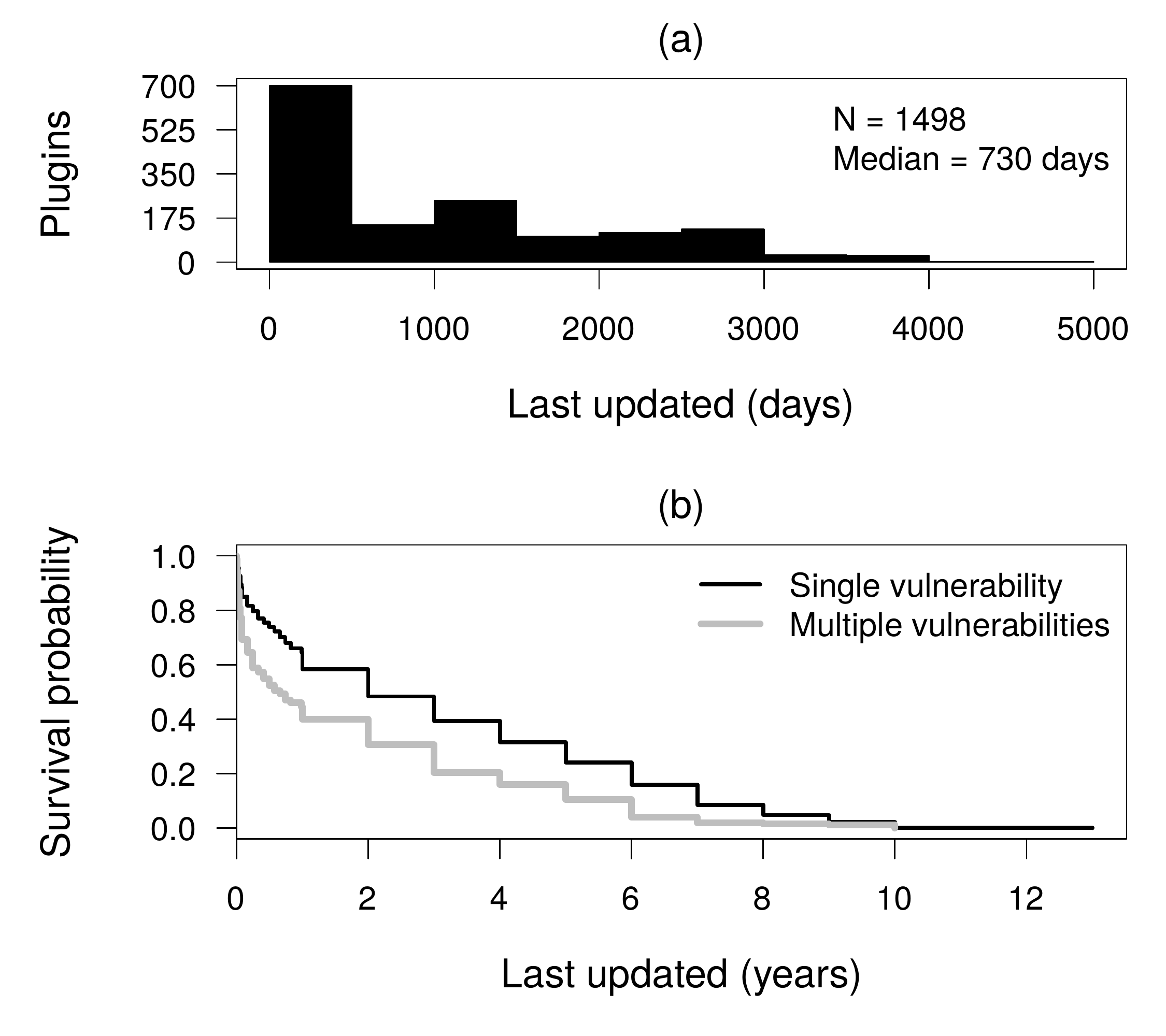}
\caption{The Most Recent Updates}
\label{fig: updates}
\end{figure}

\subsubsection{Updates}

The online portal provides also calendar time meta-data on the last updates made to the plugins. Although no documentation is provided on how the values are computed, these provide a good proxy for evaluating general maintenance effort~\cite{Hills16}. Thus: as can be seen from the plot (a) in Fig.~\ref{fig: updates}, most of the plugins have seen updates during the past two years or so. The observation is welcome because all of the plugins observed have been affected by at least one vulnerability at some point in time. However, the distribution is extremely skewed; some outlying plugins have not been updated even in a decade. These cases are sufficient to conclude that some of the plugins have been abandoned. The observation is fairly typical to large software ecosystems~\cite{Vaidya19}. Furthermore, the Kaplan-Meier survival curves~(see, e.g., \cite{Aalen08}) shown in the plot~(b) indicate that the plugins affected by only one vulnerability have been updated less frequently. The observation is logical: there is a negative correlation; when the vulnerability counts increase, the times of last updates decrease. The explanation is likely simple: bug fixes imply updates, and fixing many bugs imply frequent updates.

\subsubsection{Ratings}

Like many software portals, the WordPress plugin portal contains the common ``five-star'' rating functionality augmented with free-form comments. According to the quantitative star-ratings, most of the plugins have been reviewed positively (see Table~\ref{tab: reviews}). The standard deviation across the plugins is large, however, and there is a small tendency toward a bimodal distribution often seen with the 5-fold star-ratings \cite{Ullah16}. For the forthcoming regression analysis, a basic hypothesis is that the plugins reviewed favorably have not been affected by multiple vulnerabilities.

\begin{table}[th!b]
\centering
\caption{Review Ratings Across Plugins}
\label{tab: reviews}
\begin{tabular}{lrrrrr}
\toprule
& \multicolumn{5}{c}{Stars} \\
\hline
& One & Two & Three & Four & Five \\
\cmidrule{2-6}
Mean & 7 & 2 & 2 & 5 & 95 \\
Median & 1 & 0 & 0 & 0 & 4 \\
Standard deviation & 27 & 6 & 7 & 25 & 641 \\
\bottomrule
\end{tabular}
\end{table}

\subsubsection{Authors}

Finally, the online portal provides data about the developers of the WordPress plugins. A noteworthy observation is that only about 8\% of the plugin developers have authored multiple plugins. Consequently, a basic hypothesis is that multiple vulnerabilities have been more common for the one-shot majority.

\subsection{Regression Analysis}

\subsubsection{Setup}

The setup for the regression analysis is simple: the vulnerability counts are regressed against the meta-data variables outlined in the previous Subsection~\ref{subsec: meta-data}. Two regression models are used for the setup. By definition, the plain vulnerability counts (see Fig.~\ref{fig: counts}) are count data. Therefore, the first model estimated is a so-called ``quasi-Poisson'' regression that accounts for potential over-dispersion (that is, the variance of  counts exceeds their mean). In essence, this model yields the same coefficient estimates as the standard Poisson regression model, but a dispersion parameter $\phi$ is estimated from data and used to adjust the standard errors of the regression coefficients \cite{Zeileis08}. The second model estimated is a standard logistic regression for which the counts are truncated into dichotomous categories; the predicted values are probabilities for the plugins to be affected by multiple vulnerabilities.

\begin{table}[th!b]
\centering
\caption{Correlations (Pearson) Between Review Ratings}
\label{tab: review correlations}
\begin{tabular}{lrrrrr}
\toprule
& \multicolumn{5}{c}{Stars} \\
\hline
& One & Two & Three & Four & Five \\
\cmidrule{2-6}
One & 1.00 & 0.94 & 0.92 & 0.85 & 0.68 \\                                         
Two & 0.94 & 1.00 & 0.94 & 0.85 & 0.70 \\                                       
Three & 0.92 & 0.94 & 1.00 & 0.92 & 0.71 \\                                       
Four & 0.85 & 0.85 & 0.92 & 1.00 & 0.79 \\                                       
Five & 0.68 & 0.70 & 0.71 & 0.79 & 1.00 \\ 
\bottomrule
\end{tabular}
\end{table}

The only notable prior statistical concern with this simple regression modeling setup is about multicollinearity. Namely: the review ratings are highly correlated (see Table \ref{tab: review correlations}). As a simple solution, only the five-star ratings are included in the two models estimated. Due to the uniformly positive correlations, any of the star-ratings would suffice, however---all these yield regression coefficients with the same sign and comparable magnitudes. Hence, the statistical effect of the $5$-star ratings should be rather interpreted as an effect about whether a plugin has received any reviews to begin with.

\subsubsection{Estimates}

The results from the two regression models are summarized in Table~\ref{tab: estimates}. To ease the interpretation, the estimates from the logistic regression model are accompanied with the so-called marginal effects (MEs). These give the approximate effects directly upon the probabilities estimated (see, e.g., \cite{Ruohonen16WIMS} for details).

\begin{table}[th!b]
\centering
\caption{Regression Estimates}
\label{tab: estimates}
\begin{tabular}{lrcrr}
\toprule
& \multicolumn{1}{c}{Quasi-Poisson} && \multicolumn{2}{c}{Logistic Regression} \\
\cmidrule{2-2}\cmidrule{4-5}
& Coefficient && Coefficient & ME \\
\hline
\texttt{(Intercept)} 
& \phantom{-}0.069$^{\phantom{***}}$ && 
-2.357$^{***}$ & -- \\
\texttt{Deprecated}
& \phantom{-}0.170$^{**\phantom{*}}$ && 
0.560$^{**\phantom{*}}$ & \phantom{-}0.100 \\
\texttt{LastUpdated}
& -0.036$^{**\phantom{*}}$ &&
-0.092$^{*\phantom{**}}$ & -0.016 \\
\texttt{FiveStars}
& $<$0.001$^{***}$ && 
$<$0.001$^{\phantom{***}}$ & $<$0.001 \\
\texttt{OneShotAuthor}
& \phantom{-}0.184$^{*\phantom{**}}$ && 
\phantom{-}0.341$^{\phantom{***}}$ & \phantom{-}0.057 \\
\texttt{$[100, 1000)$}
& \phantom{-}0.074$^{\phantom{***}}$ && 
\phantom{-}0.644$^{**\phantom{*}}$ & \phantom{-}0.117 \\
\texttt{$[1000, 10000)$}
& \phantom{-}0.203$^{**\phantom{*}}$ && 
\phantom{-}1.118$^{***}$ & \phantom{-}0.205 \\
\texttt{$[10000, 100000)$}
& \phantom{-}0.369$^{***}$ &&
\phantom{-}1.610$^{***}$ & \phantom{-}0.313 \\
\texttt{$[100000, \infty)$}
& \phantom{-}0.849$^{***}$ &&
\phantom{-}2.131$^{***}$ & \phantom{-}0.445 \\
\bottomrule
\multicolumn{5}{l}{\small $\textmd{N} = 1498$, $\hat{\phi} = 0.98$, $^{***}~\textmd{for}~p < 0.001$, $^{**}~\textmd{for}~p < 0.01$, $^{*}~\textmd{for}~p < 0.05$.} \\
\end{tabular}
\end{table}

In general, the two models agree well with each other; the signs of the coefficients are consistent, for instance. The estimated dispersion parameter for the quasi-Poisson model indicates no particular concern about over-dispersion. For unpacking the effects of the individual variables, it can be started by noting that  deprecated plugins tend to increase vulnerability counts. The observation seems logical. The maintainers of the online portal may deprecate plugins with many unfixed vulnerabilities, for instance. As was expected, increasing lags in the update times tend to decrease the vulnerability counts; fixing multiple vulnerabilities requires more frequent updates. The effect of the $5$-star review ratings is positive but negligible in magnitude. This observation supports earlier results~\cite{Koskinen12}. Likewise, the effect of one-shot plugin authors is also positive but small. In contrast, the magnitudes are large for all of the re-coded dummy variables approximating the installation amounts. For instance: when compared to plugins with less than a hundred installations (cf.~Fig.~\ref{fig: installations}), the plugins with more than one hundred thousand online deployments have about $0.445$ higher probability of being affected by multiple vulnerabilities, all other things being constant. Consequently, the demand-side viewpoint seems to hold.

\subsubsection{Diagnostics}

The logistic regression model can be taken under a brief further inspection. To begin with, it should be remarked that the overall performance is modest. For instance, the so-called area under the curve (AUC) in a receiver operating characteristic curve is $0.711$. Most of the performance is attributable to the installation amounts. When only these are included, $\textmd{AUC} = 0.694$.

\begin{figure}[th!b]
\centering
\includegraphics[width=\linewidth, height=3.5cm]{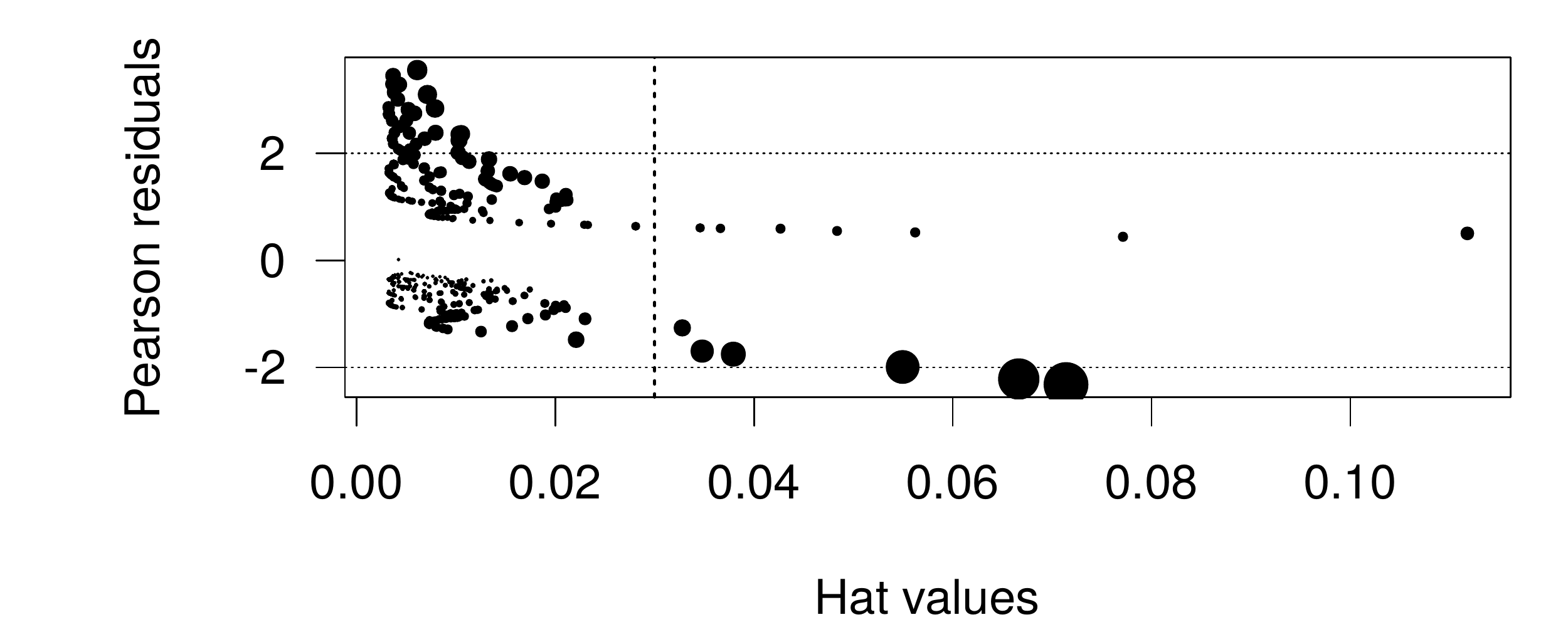}
\caption{Influence Plot (Fox-Weisberg)}
\label{fig: influence}
\end{figure}

When the so-called Pearson residuals are examined, it is evident that there are large residuals in the logistic regression model estimated. However, Fox's test for outliers \cite{Fox16}, as implemented in the \texttt{outlierTest} function for the \textit{car} package~\cite{FoxWeisberg11}, indicates only one outlier, and no outliers when a Bonferonni correction is applied. This test result does not rule out the potential for particularly influential observations that may change the coefficient estimates. To examine such influential observations, a so-called influence plot provides a good graphical diagnostic tool. It plots the Pearson residuals against the so-called hat values, and further scales the areas of the plotted observations according to Cook's distance~\cite{FoxWeisberg11}. Without delving into the statistical details (see \cite{Santner89} for a good take on the mathematical background), the resulting influence plot is shown in Fig.~\ref{fig: influence}. If the thirteen observations on the right-hand side of the dotted vertical line are removed and the logistic regression model is re-estimated with the reduced dataset, the coefficient estimates are highly similar to those in Table~\ref{tab: estimates}. The marginal effects of the installation amounts are $0.121$, $0.209$, $0.320$, and $0.440$. For any practical purposes, the estimates are equal. The same conclusion is reached with analogous omissions according to the large Pearson residuals.

\subsubsection{Confounding factors}

A more fundamental question is whether there are confounding factors or omitted variables that should be taken into account. While it is clear that XSS in particular is statistically associated with the vulnerability counts, it may also be that some particular weakness types interfere with the independent metrics used for modeling. For instance, the last updates made to the plugins (see Fig.~\ref{fig: updates}) might be assumed to vary according to the CWEs in Fig.~\ref{fig: cvss cwe}. Such assumptions are not easy to examine, however. As was discussed in Section~\ref{sec: data}, the regression analysis operates at the plugin-level, which makes it difficult to incorporate vulnerability-level metrics. The abstraction inconsistencies between WPVDB and NVD cause additional problems. External validity issues would be also introduced due to the small amount of CVE-referenced vulnerabilities in the sample. A further question is whether inference with CWEs is theoretically sensible in the present context. For instance, previous results indicate that most security bug fixes for WordPress plugins require changing only a few lines of code, although even such small changes take a long time to implement by the plugin developers~\cite{Mesa18}. The reasons for such results may not necessarily relate to the technical details about the vulnerabilities themselves, but perhaps more to the general code quality and effort devoted to maintaining WordPress plugins. Against this backdrop, it may be that more plausible confounding factors would be available by examining code-level and other metrics traditionally used in empirical software engineering.

\section{Discussion}\label{sec: discussion}

\subsection{Conclusion}

This paper examined a so-called demand-side viewpoint to vulnerabilities in WordPress plugins. The underlying rationale behind the viewpoint seems sensible according to the empirical results. In other words, the answer to the RQ is positive: widespread adoption and large installation bases are statistically associated with larger vulnerability amounts. If installation bases provide an important incentive on the black-hat side~\cite{Moore16}, these seem to provide an incentive also on the white-hat side. In other words, there is only a small incentive to devote time and effort to discover, document, and disclose vulnerabilities from a ``Joe's basketball plugin''. Needless to say, incentives and the associated supply and demand factors do not necessarily tell anything about \textit{actual} security. Given the abundance of static analysis tools for PHP code~\cite{Koskinen12, Nunes18, Nunes19, Santos17}, a~more rigorous code-level validation of the demand-side viewpoint would also offer one plausible approach for further empirical research. Static analysis and more generally code-level assays would allow to also better understand the apparent maintenance issues.

\subsection{Limitations}

A notable limitation relates to the reliability of the dataset assembled. In particular, the reliability of WPVDB's vulnerability data has been debated~\cite{PluginVulnerabilities18a}. Even though no research has been done to examine these debates in detail, empirical reliability issues should be still acknowledged as a potential limitation. After all, even NVD has been shown to occasionally contain some inaccuracies~\cite{NguyenMassacci15}. An analogous concern applies to the meta-data scraped from the WordPress plugin portal. That said, these potential reliability problems should not be exaggerated. Some assurance is available by noting that WPVDB's vulnerability data has been used in previous research~\cite{Nunes18, Nunes19}. The same goes for the meta-data from the online WordPress plugin portal~\cite{Hills16, Koskinen12}. Because the demand-side viewpoint pursued does not relate to security \textit{per~se}, some inaccuracies can be also accepted. In a similar vein: vulnerability counts should not be used to judge the security of a software product~\cite{Ruohonen18IWESEP, Walden10a}, but these are appropriate for analyzing incentives to find vulnerabilities.

\subsection{Related Work}

Different software ecosystems have received a great deal of attention in recent years. While there are many reasons for the attention, one fundamental reason is the recent explosion of dependencies between libraries and related artifacts. This dependency explosion has also intensified the age-old relation between maintenance and security. From a practical maintenance viewpoint, it ``sounds incredibly unsafe'' to trust code downloaded from ``a stranger on the internet''---why ``would anyone do this?'' \cite{Cox19}. While the answers to the question are still unclear, good progress has been made to better understand vulnerabilities within software ecosystems~\cite{ZapataKula18, Ruohonen18IWESEP}. Recently, the issues examined have been further extended toward the security of ecosystems themselves~\cite{Vaidya19}. Although the dependency mechanisms are different, also the WordPress plugins examined can be placed into this ecosystem context. In terms of WordPress plugins, previous work has been done to address the dependency mechanisms~\cite{Hills16}, the known vulnerabilities ~\cite{Mesa18, Walden10b}, the testing of these~\cite{Nunes19}, and the exploits for these~\cite{Trunde15}. There is also at least one study that has addressed the meta-data aspects such as user reviews~\cite{Koskinen12}. However, neither these previous works nor this paper explicitly address a question about whether anything can be done to help those downloading code from strangers on the Internet.

\subsection{Toward Recommendation Systems}

The empirical results presented can be further portrayed from a different angle. There has been an increasing interest to examine and develop different recommendation systems for OSS libraries~\cite{Kula18a}. Security is one aspect to consider when choosing a library, plugin, or other complementary artifact for a software project. To this end, also vulnerability-based metrics have been proposed. For instance, some have defined metrics with the goal of providing tools for ``selecting better versions of OSS, where definition of \textit{better} is fewer vulnerabilities''~\cite{ZhangMalhotra18}. If the installation amounts implicitly reflect quality and better software in general, a practical recommendation might in fact be exactly the opposite: it may be preferable to pick a WordPress plugin with a large installation base, which tends to result in \textit{more} vulnerabilities. If also CVEs are allocated for the vulnerabilities, there is a good chance that the given plugin is relatively well-maintained. Given the background of security folklore, also this tentative recommendation can be seen as counterintuitive. Further security-related considerations include vulnerability density in terms of software size~\cite{Huynh10, Koskinen12}, the availability of security documentation and associated resources~\cite{Walden10a}, and many related commonsense aspects~\cite{Cox19}. Of course, it may also be that security is not a factor in adoption decisions, as hinted by a recent industry study~\cite{Pano18}. Though, many of the decision factors reported in the noted study kind of rob Peter to pay Paul: code complexity, size of an open source community and its responsiveness, and many related factors presumably correlate with vulnerability~counts.

%\balance
\bibliographystyle{abbrv}
%\bibliography{vuln}

\end{document}